\documentclass[onecolumn, 10p, journalt]{IEEEtran}
\usepackage{amsmath,amssymb,eucal,graphicx,wrapfig}%,doublespace}
\usepackage{pst-plot}
\usepackage{pst-node}
\usepackage{epsfig}
\usepackage{amsthm}
\usepackage{amsfonts}
\usepackage{epsfig}
\usepackage{float}
\usepackage{pstricks}
\usepackage{color}
\usepackage{multicol}
\usepackage{subfigure,url}

\newtheorem{thm}{Theorem}

\theoremstyle{remark}

\theoremstyle{definition}

\begin{document}
\title{A Lattice Compress-and-Forward Scheme}
\author{Yiwei Song and Natasha Devroye \\
Department of Electrical and Computer Engineering \\
University of Illinois at Chicago \\
{\tt ysong34, devroye@ ece.uic.edu}
}

% make the title area
\maketitle

\centerline{\today}

\bigskip

\begin{abstract}
We present a nested lattice-code-based strategy that achieves the random-coding based Compress-and-Forward (CF) rate for the three node Gaussian relay channel. To do so, we first outline a lattice-based strategy for the $(X+Z_1,X+Z_2)$ Wyner-Ziv lossy source-coding with side-information problem in Gaussian noise, a re-interpretation of the nested lattice-code-based  Gaussian Wyner-Ziv scheme presented by Zamir, Shamai, and Erez.  We use the notation $(X+Z_1,X+Z_2)$ Wyner-Ziv to mean that the source is of the form $X+ Z_1$ and the side-information at the receiver is of the form $X+ Z_2$, for independent Gaussian $X, Z_1$ and $Z_2$.  We next use this $(X+Z_1,X+Z_2)$  Wyner-Ziv scheme to implement a ``structured'' or lattice-code-based CF scheme which achieves the classic  CF rate for Gaussian relay channels. This suggests that lattice codes may not only be useful in point-to-point single-hop source and channel coding, in multiple access and broadcast channels, but that they may also be useful in larger relay networks. The usage of lattice codes in larger networks is motivated by their structured nature (possibly leading to rate gains) and decoding (relatively simple) being more practically realizable than their random coding based counterparts. We furthermore expect the proposed lattice-based CF scheme to constitute a first step towards a generic structured 
achievability scheme for networks such as a structured version of  the recently introduced ``noisy network coding''. 
%which, in the long term, we e where we expect a structured version of the recent noisy network coding which generalized the CF scheme, to emerge\nrd{How about focusing on 'practical way to implement lattice' and 'structured noisy network coding'}.
\end{abstract}

\section{Introduction}

Lattice codes have been shown to perform as well as classic Shannon random codes for certain Gaussian channels, and to outperform random codes for specific Gaussian channels. This leads to the general question of whether the performance of random codes in Gaussian channels may be approached or even exceeded using carefully designed lattice codes. As much is known about lattice codes and their performance in simple point-to-point source and channel coding scenarios, in this paper we take the next step towards the goal of demonstrating that lattices may mimic random codes in Gaussian networks and consider the simple three user Gaussian relay channel.  In \cite{Song:2010:latticerelay} it was shown that lattice codes may achieve the Gaussian Decode-and-Forward rate of \cite{Cover:1979:relay} for the Gaussian relay channel. We now demonstrate that lattice codes may also be used to achieve the Gaussian Compress-and-Forward (CF) rate of \cite{Cover:1979:relay} for this channel.

{\bf Scenarios in which lattice codes achieve the same rates as random codes.} Lattice codes (and lattice decoding) have been shown to be capacity achieving in the Additive White Gaussian Noise (AWGN) point-to-point channel, using a unique decoding  technique \cite{Erez:2004} exploiting a carefully chosen Minimum Mean Squared Error (MMSE) scaling coefficient,  and recently,  using an alternative list decoding technique \cite{Song:2010:latticerelay}. Lattices codes may also be constructed that achieve the capacity of the Gaussian Multiple Access Channel (MAC) \cite{Nazer:2009:computeandforward} and the Gaussian Broadcast Channel (BC) \cite{Zamir:2002:binning}. The latter exploited the fact that lattice codes may achieve the dirty-paper coding channel capacity \cite{Zamir:2002:binning} by mimicing random binning techniques in a structured manner.
Recently, using a lattice list-decoding technique, nested lattice codes were shown to achieve the Gaussian random coding Decode-and-Forward rate in the Gaussian relay channel \cite{Song:2010:latticerelay}.
%whether lattices are able to mimic random codes in Compress-and-Forward scenarios remained an open problem which we settle here.

{\bf Scenarios in which lattice codes may outperform random codes.} Lattice codes provide structured codebooks. Intuitively, this may be exploited to achieve higher rates than unstructured or random codebooks, particularly in scenarios where combinations of codewords are decoded. Decoding the ``sum'' of codewords may be done at a higher rate by structured codes than random codes as the ``sum'' of two structured codewords may be designed to again be a codeword, whereas the sum of two random codewords is with high probability not another codeword. In the latter, decoding the sum of two codewords is equivalent to decoding them individually, leading to more stringent rate constraints than if we are simply able to ``decode the sum'' (and not be forced to decode the individuals) using structured codes. This property is exploited in the  compute-and-forward framework \cite{Nazer:2009:computeandforward}, in which various linear combinations of messages are decoded, as well as in the  two-way relay channel without direct links \cite{1,2}, and with direct links \cite{Song:2010:latticerelay} to achieve higher rates than those known to be achievable with random codebooks. Finally, this property has been exploited in several $K>2$ user interference channels to decode the sum of interference terms \cite{sridharan_lattice}.

{\bf Lattice codes for binning.} In networks with side-information, the concept of binning, which effectively allows the transmitters and receivers to properly exploit this side-information, is critical. The usage of lattices and structured codes for binning (as opposed to random binning as previously proposed) in various types of networks was considered in a comprehensive fashion  in \cite{Zamir:2002:binning}. Of particular interest to the problem considered here is the nested lattice-coding approach of \cite{Zamir:2002:binning} to the Gaussian Wyner-Ziv coding problem. The Wyner-Ziv coding problem is that of lossy source coding with correlated side-information at the receiver or reconstructing node. One example of a Gaussian Wyner-Ziv problem is one in which the Gaussian source to be compressed is of the form $X+Z$, and the side-information available at the reconstructing node is $X$, for $Z$ independent of $X$ and Gaussian, which we term the $(X+Z,X)$ Wyner-Ziv problem\footnote{More generally, the source to be compressed is $X$ with correlated side-information $Y$ at the receiver.}. A lattice-scheme is provided in \cite{Zamir:2002:binning} for the $(X+Z,X)$ Wyner-Ziv problem. We consider a lattice Wyner-Ziv coding scheme for the slightly altered $(X+Z_1,X+Z_2)$ channel model in which the source to be compressed is of the form $X+ Z_1$ and the side-information is of the form $X+ Z_2$, for independent, Gaussian $X,Z_1$ and $Z_2$. We present a lattice-scheme for the $(X+Z_1,X+Z_2)$ Wyner-Ziv problem, which may be viewed as a re-interpretation of the scheme of \cite{Zamir:2002:binning} for the $(X+Z,X)$ model (which was only mentioned in a footnote and not fully presented in \cite{Zamir:2002:binning}). We include this lattice-based scheme for the $(X+Z_1,X+Z_2)$ model for completeness, as it will be used in our main result on a lattice CF scheme. We provide additional insight into the relationship between the $(X+Z,X)$ and $(X+Z_1,X+Z_2)$ models, and use the latter to construct a CF scheme based on nested lattice codes which recovers the same achievable rate as the classic achievable CF rate \cite{Cover:1979:relay} for the Gaussian relay channel.
%Then we apply this lattice approach of Wyner-Ziv coding to Compress-and-Forward scheme in three nodes relay channel,  and it fully recovers the achievable rate region of CF.\\

{\bf The classic Compress-and-Forward (CF) rate for the Gaussian relay channel.} Cover and El Gamal first proposed a CF scheme for the three user relay channel in \cite{Cover:1979:relay}. In it, the relay does not decode the message (as it would in the Decode-and-Forward scheme) but instead compresses its received signal and forwards the compression index. The destination first recovers the compressed signal, using its direct-link  side-information (the Wyner-Ziv problem), and then proceeds to decode the message from the recovered  compressed signal. The CF  scheme is generalized to arbitrary relay networks in the recently proposed ``noisy network coding'' scheme \cite{lim2010noisy}. 
%\nrd{YIWEI: is this all you want to say about noisy network coding? Perhaps another sentence? Maybe we should, but have not got a good one yet}. 
Armed with a lattice Wyner-Ziv scheme, we mimic every step of the classic CF scheme using lattice codes and will show that the same rate may be achieved in a structured manner.

%\nrd{Nice introduction and summary :)} \\
{\bf Contribution and paper organization.} The central contribution of this work is the application of a general lattice-coding based Wyner-Ziv scheme to the Gaussian three node relay channel. In particular, in Section \ref{sec:notation} we first outline our notation and nested lattice coding preliminaries. In Section \ref{sec:WZ} we outline a nested lattice-code based scheme for a $(X+Z_1,X+Z_2)$ Wyner-Ziv problem in Theorem \ref{thm:WZ}, providing an in-depth look at the scheme mentioned in footnote 6 of \cite{Zamir:2002:binning}. Using the scheme of Section \ref{sec:WZ}, in Section \ref{sec:LCF}, in Theorem \ref{thm:LCF},  we show that the rate achieved by random codes in the classic Compress-and-Forward scheme may be achieved using nested lattice codes. Finally, we conclude in Section \ref{sec:conclusion}. Given the structure of lattice codes, this may constitute a more practical implementation of Wyner-Ziv coding (as already noted in \cite{Zamir:2002:binning}),  of the CF scheme, and is an important first step towards a generic ``structured'' achievability scheme for networks such as a ``structured'' noisy network coding \cite{lim2010noisy}.

\section{Preliminaries a nested lattice codes}
\label{sec:notation}
We first outline our notation and definitions for nested lattice codes for transmission over AWGN channels, following those of \cite{Zamir:2002:binning, nam:2009nested}. We note that  \cite{loeliger1997averaging, Zamir:2002:binning,  Erez:2004} and in particular \cite{zamir-lattices} offer more thorough treatments, and defer the interested reader to those works for more details.  An $n$-dimensional lattice $\Lambda$ is a discrete subgroup of Euclidean space $\mathbb{R}^n$ (of vectors ${\bf x}$, though we will denote these without the bold font as $x$) with Euclidean norm $|| \cdot ||$ under vector addition. We may define
% and may be expressed as all integral combinations of basis vectors ${\bf g_i}\in {\mathbb R}^n$
%\[ \Lambda = \{ \lambda = G \; {\bf i}: \; {\bf i}\in \mathbb{Z}^n\},\]
%for $\mathbb{Z}$ the set of integers, and $G := [{\bf g_1} | {\bf g_2}| \cdots {\bf g_n}]$ the $n\times n$ generator matrix corresponding to the lattice $\Lambda$. Further define:

$\bullet$ The {\it nearest neighbor lattice quantizer} of $\Lambda$ as $ Q_\Lambda({x}) = \arg \min_{\lambda\in \Lambda} ||{x}-\lambda||;$

$\bullet$ The { \texttt{mod }$\Lambda$} operation as ${x}$  \texttt{mod }$\Lambda : = {x} - Q_\Lambda({x})$, hence
$ {x} = Q_{\Lambda}({x}) + ({x} \mod \Lambda);$

$\bullet$ The {\it fundamental region of $\Lambda$} as the set of all points closer to the origin than to any other lattice point $\mathcal{V}(\Lambda):= \{{x}:Q({x}) = {\bf 0}\}$
which is of volume $V: = \mbox{Vol}({\mathcal V}(\Lambda))$;

$\bullet$ The {\it second moment per dimension of a uniform distribution over ${\mathcal V}$} as
$ \sigma^2(\Lambda) : = \frac{1}{V}\cdot \frac{1}{n} \int_{\mathcal V} ||{x}||^2 \; d{x}$;

$\bullet$ The {\it Crypto lemma} \cite{Forney:ShannonWinener:2003} which states that  $({x} + {U}) \mod \Lambda$ (where ${U}$ is uniformly distributed over $\mathcal{V}$) is an independent random variable uniformly distributed over $\mathcal{V}$.

Standard definitions of {\it Poltyrev good} and {\it Rogers good} lattices are used \cite{Erez:2004}, and by \cite{Erez:latticegood:2005} we are guaranteed the existence of lattices which are both Polytrev and Rogers good, which may intuitively be thought of as being good channel and source codes, respectively.

The proposed schemes will be based on {\it nested lattice codes}. To define these, consider two lattices $\Lambda$ and $\Lambda_c$ such that $\Lambda \subseteq \Lambda_c$ with fundamental regions ${\cal V}, {\cal V}_c$ of volumes $V, V_c$ (where $V\geq V_c$) respectively.
Here $\Lambda$ is called the {\it coarse} lattice which is a sublattice of  $\Lambda_c$,  the {\it fine} lattice.  We denote the cardinality of a set $A$ by $|A|$. The set  $ \mathcal{C}_{\Lambda_c, {\cal V}} = \{ \Lambda_c \cap \mathcal{V} \} $ may be employed as the codebook for transmission over the AWGN channel, with coding rate $R$ defined as
\[ R = \frac{1}{n} \log |\mathcal{C}_{\Lambda_c, {\cal V}}| = \frac{1}{n} \log \frac{V}{V_c}.\] Here $\rho = |\mathcal{C}_{\Lambda_c, {\cal V}}|^{\frac{1}{n}} = \left( \frac{V}{V_c} \right)^{\frac{1}{n}}$ is the nesting ratio of  this {\it nested  $(\Lambda, \Lambda_c)$ lattice code pair}. A pair of {\it good} nested lattice codes, where $\Lambda$ is both {\it Rogers good} and {\it Poltyrev good}  and $\Lambda_c$ is {\it Poltyrev good}, were shown to exist and be capacity achieving (as $n\rightarrow \infty$) for the AWGN channel \cite{Erez:2004}. The {\it goodness} of lattice code pairs may be extended to a nested lattice chain, which consists of nested lattice codes  $\Lambda \subseteq \Lambda_1 \subseteq \Lambda_2$ which may be {\it Rogers good} and {\it Poltyrev good} for arbitrary nesting ratios \cite{Krithivasan:2007:goodlattice} . 
%\nrd{nice summary..:)}
%{\it Good} lattice chains will be used in the achievability schemes for both models, described next.

\section{Lattice codes for the $(X+Z_1,X+Z_2)$ Wyner-Ziv model}
\label{sec:WZ}

{\bf Problem statement.} We consider the lossy compression of the Gaussian source $Y = X+Z_1$, with side-information $X+Z_2$ available at the reconstruction node, where $X, Z_1$ and $Z_2$ are independent zero mean Gaussian random variables of variance $P, N_1$, and $N_2$ respectively. We note that, with slight abuse of notation, $X, Z_1$ and $Z_2$ denote $n$-dimensional vectors where $n$ is the classic blocklength, or number of channel uses, which will tend to infinity.  The rate-distortion function for the source $X+Z_1$ taking on values in ${\cal X}_1$ with side-information $X+Z_2$ taking on values in ${\cal X}_2$ is defined as the minimum rate required to achieve a distortion $D$ when $X+Z_2$ is available at the decoder. To be more specific, it is the infimum of rates $R$ such that there exist maps $i_n: {\cal X}_1  \rightarrow \{1,2,\cdots, 2^{nR}\}$ and $g_n: {\cal X}_2 \times \{1,2, \cdots, 2^{nR}\}\rightarrow {\cal X}_1$ such that $\lim \sup_{n\rightarrow \infty} E[d(X+Z_1 , g_n(X+Z_2, i_n(X+Z_1))]\leq D$ for some distortion measure $d(\cdot, \cdot)$.
%The encoding and decoding functions take the forms
%\[  \mbox{encoded index } i = f(X+Z_1), \;\;\; \mbox{ reconstructed source } \widehat{Y} = \widehat{X+Z_1} = g(i,X+Z_2) \]
%for some functions $f(\cdot)$ and $g(\cdot, \cdot)$. The source coding rate corresponds to $\frac{1}{n}\log(\mbox{the cardinality of the range of }f(\cdot))$. In lossy source coding, the smallest rate $R(D)$ at which one may reconstruct the source to a given expected distortion $\frac{1}{n} E[d(X,\widehat{X})]\leq D$ is of interest.
 If the distortion measure $d(\cdot, \cdot)$ is the squared error distortion, $d(X,\widehat{X}) = \frac{1}{n}E[||X-\widehat{X}||^2]$, then, by  \cite{wyner1978rate},
  the rate distortion function $R(D)$ for the source $X+Z_1$ given the side-information $X+Z_2$ is given by
\begin{align*}
R(D) &= \frac{1}{2} \log \left(\frac{\sigma^2_{X+Z_1|X+Z_2}}{D} \right),  \qquad 0 \leq D \leq \sigma^2_{X + Z_1|X+Z_2}\\
&= \frac{1}{2} \log \left( \frac{N_1 + \frac{PN_2}{P+N_2}}{D} \right),  \qquad 0 \leq D \leq N_1 + \frac{PN_2}{P+N_2},
\end{align*}
and $0$ otherwise,  where $\sigma^2_{X+Z_1|X+Z_2}$ is the conditional variance of $X+Z_1$ given $X+Z_2$.
 We note that the lattice-code implementation of the Wyner-Ziv scheme of \cite{Zamir:2002:binning} considered the lossy compression of the source $X+Z$ with side-information $X$ at the reconstruction node. In footnote 6 on pg. 1260 of \cite{Zamir:2002:binning} it is stated that this model may WLOG be used to capture all general jointly Gaussian sources and side-informations (including the aforementioned source $X+Z_1$ with side-information $X+Z_2$). 
 %, though no details are provided. 
 The scheme we present next is an example of this more general scheme, and is provided only for completeness; the scheme is essentially identical to that of \cite{Zamir:2002:binning}, with a few careful adjustments made. We use the  lattice-based scheme presented next to derive a lattice Compress-and-Forward scheme in Section \ref{sec:LCF}.
 %mentioned the generall form of source and side information in the footnote, it does not provide the details. Here, we describe the implement details of this case and use it in lattice CF in next section.

\begin{figure}[ht]
\centering
\includegraphics[width=18cm]{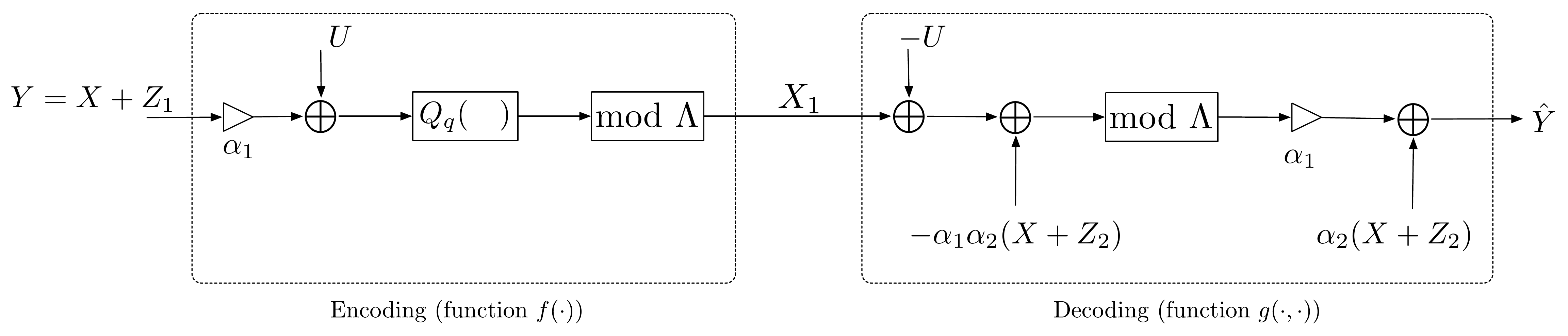}
\caption{Lattice coding for the $(X+Z_1,X+Z_2)$ Wyner-Ziv problem.}
\label{fig:WZ}
\end{figure}

\begin{thm}
The following rate-distortion function for the lossy compression of the source $X+Z_1$ subject to the reconstruction side-information $X+Z_2$ and squared error distortion metric may be achieved through the use of lattice codes:
\[ R(D)
= \frac{1}{2} \log \left( \frac{N_1 + \frac{PN_2}{P+N_2}}{D} \right),  \qquad 0 \leq D \leq N_1 + \frac{PN_2}{P+N_2},\]
and $0$ otherwise.
\label{thm:WZ}
\end{thm}

The remainder of this Section consists of the proof of Theorem \ref{thm:WZ}.

\medskip
{\bf General lattice Wyner-Ziv.} Consider a pair of nested lattice codes $\Lambda \subseteq \Lambda_q$, where $\Lambda$ is Poltyrev-good with second moment $N_1 + \frac{PN_2}{P + N_2}$, and $\Lambda_q$ is Rogers-good with second moment $D$. We consider the encoding and decoding schemes of Fig. \ref{fig:WZ}. We let $U$ be a quantization dither signal which is uniformly distributed over $\mathcal{V}(\Lambda_q)$, and introduce  the following MMSE coefficients, whose choices will be justified later:
\begin{equation}
 \alpha_1  = \sqrt{1 - \frac{D}{N_1 + \frac{PN_2}{P+N_2}}}, \;\;\;\; \alpha_2= \frac{P}{P+ N_2}.
 \label{eq:alpha}
 \end{equation}
%be the MMSE coefficients (we will justify these choices later on).

{\bf Encoding.} The encoder quantizes the scaled and dithered signal $\alpha_1(X+Z_1)+U$ to the nearest fine lattice point, which is then modulo-ed back to the Voronoi region of coarse lattice as
\begin{align*}
I &= Q_q( \alpha_1 (X+Z_1) + U ) \mod \Lambda \\
&= ( \alpha_1 (X+Z_1) + U - ( \alpha_1(X+Z_1) + U) \mod \Lambda_q ) \mod \Lambda \\
&= ( \alpha_1 (X+Z_1) + U - E_q ) \mod \Lambda,
\end{align*}

where $E_q := ( \alpha_1(X+Z_1) + U) \mod \Lambda_q$ is independent of everything else and uniformly distributed over $\mathcal{V}(\Lambda_q)$ according to the Crypto lemma \cite{Forney:ShannonWinener:2003}.  The encoder sends the index $i$ of $I$ to the decoder at the source coding rate
\begin{align*}
R &= \frac{1}{n} \log \left( \frac{V(\Lambda)}{V(\Lambda_q)} \right) = \frac{1}{2} \log \left( \frac{\sigma^2(\Lambda)}{\sigma^2(\Lambda_q)} \right) = \frac{1}{2} \log \left( \frac{N_1 + \frac{PN_2}{P+N_2}}{D} \right).
\end{align*}

{\bf Decoding.} The decoder receives the index $i$ of $I$ and reconstructs $\widehat{Y}$ as
\begin{align*}
\widehat{Y} &=  \alpha_1( (I - U - \alpha_1 \alpha_2 (X+Z_2) ) \mod \Lambda ) + \alpha_2( X + Z_2) \\
&= \alpha_1 ( ( \alpha_1 ( X+ Z_1) + U - E_q - U - \alpha_1\alpha_2(X+Z_2) )\mod \Lambda  ) + \alpha_2(X+Z_2) \\
&= \alpha_1 ( ( \alpha_1( (1-\alpha_2) X - \alpha_2 Z_2 + Z_1) - E_q ) \mod \Lambda ) + \alpha_2( X + Z_2)  \\
&\equiv \alpha_1 ( \alpha_1( (1 - \alpha_2) X - \alpha_2 Z_2 + Z_1) - E_q )  + \alpha_2(X+ Z_2) \\
& = ( \alpha_1^2 - \alpha_1^2\alpha_2 + \alpha_2) X + \alpha_2 ( 1 - \alpha_1^2) Z_2 + \alpha_1^2 Z_1 - \alpha_1 E_q,
\end{align*}
where the fourth equivalence is meant to denote asymptotic equivalence (as $n\rightarrow \infty$), since, as in  \cite{Zamir:2002:binning}
\begin{align}
Pr \{ ( \alpha_1( (1-\alpha_2) X - \alpha_2 Z_2 + Z_1) - E_q ) \mod \Lambda \neq  \alpha_1( (1-\alpha_2) X - \alpha_2 Z_2 + Z_1) - E_q  \} \label{pe}
 \end{align}
goes to $0$ as $n \rightarrow \infty$ for a sequence of a good nested lattice codes since
\begin{align}
\frac{1}{n} &E || \alpha_1( (1-\alpha_2) X - \alpha_2 Z_2 + Z_1) - E_q ||^2 = \alpha_1^2 \left( \frac{PN_2}{P + N_2} + N_1 \right) + D = \frac{PN_2}{P+ N_2} + N_1 = \sigma^2(\Lambda).\label{eq:imp}
\end{align}
The careful choice of the MMSE coefficients $\alpha_1$ and $\alpha_2$ as in \eqref{eq:alpha} is reflected so as to guarantee the above equation \eqref{eq:imp}.  Thus,
\begin{align*}
\widehat{Y} - Y &= ( \alpha_1^2 - \alpha_1^2\alpha_2 + \alpha_2) X + \alpha_2 ( 1 - \alpha_1^2) Z_2 + \alpha_1^2 Z_1 - \alpha_1 E_q  - (X+Z_1) \\
&= -(1-\alpha_1^2)(1-\alpha_2)X + \alpha_2(1-\alpha_1^2)Z_2 - (1-\alpha_1^2)Z_1 -  \alpha_1 E_q \\
&= - (1 - \alpha_1^2) ( (1 - \alpha_2) X - \alpha_2 Z_2 + Z_1) - \alpha_1 E_q,
\end{align*}
from which we may bound the squared error distortion as
\begin{align*}
\frac{1}{n} E || \widehat{Y} - Y ||^2 &=  ( 1 - \alpha_1^2)^2 \left( \frac{PN_2}{P + N_2} + N_1\right) + \alpha_1^2 D = D,
\end{align*}
again through the careful choice of $\alpha_1$ and $\alpha_2$ as in \eqref{eq:alpha}.
%\nrd{YIWEI: is everything with equality for the distortion, never $\leq$? Seems strange to have exact equality.... I thought about it too. But Zamir's paper use the equality everywhere and it is a little messy to change to $\leq$. So we are showing the extreme case.}

\medskip
{\bf Remarks on the MMSE coefficients $\alpha_1$ and $\alpha_2$.}
We first note that the source
%$\alpha_2$ is equivalent to $a$ in the footnote 6 on page 1260 of \cite{Zamir:2002:binning} as the source
$X+Z_1$ may be expressed as
\[ X+ Z_1 = \alpha_2(X+Z_2)  + (1-\alpha_2) X + Z_1 - \alpha_2 Z_2, \]
and that by choosing $\alpha_2 = \frac{P}{P+N_2} $, $X+Z_2$ and $(1-\alpha_2) X + Z_1 - \alpha_2 Z_2$ are independent since
\begin{align*}
E[ (X+ Z_2)((1-\alpha_2) X + Z_1 - \alpha_2 Z_2)] &= (1-\alpha_2) E(X^2) - \alpha_2 E(Z_2^2)= 0.
\end{align*}
In this case, we are able to equate $\alpha_2$ with the $a$ of footnote 6 on pg. 1260 of \cite{Zamir:2002:binning}, thereby relating the above scheme to that of \cite{Zamir:2002:binning}.  In this case, we may intuitively think of  $\alpha_1$ as a source coding MMSE coefficient, and of  $\alpha_2$ as a channel coding MMSE coefficient,  since it plays a role similar to the MMSE coefficient used in the lattice channel coding problem \cite{Erez:2004}, i.e. it minimizes $E[(1-\alpha_2)X - \alpha_2 Z_2 ]^2$. In particular, we may see the importance of the correct choice of these coefficients by considering the alternative choices of $\alpha_1$ and $\alpha_2$, with the corresponding suboptimal rates:
\begin{itemize}
\item If $\alpha_1$ is set to 1 (which means we actually do not use it),  and the second moment of the coarse lattice is changed accordingly, the rate distortion function is
\[ R(D) =  \frac{1}{2} \log \left(1 + \frac{N_1 + \frac{PN_2}{P+N_2}}{D} \right) > \frac{1}{2} \log \left( \frac{N_1 + \frac{PN_2}{P+N_2}}{D} \right). \]
\item If $\alpha_2$ is set to 1, and the second moment of the coarse lattice is changed accordingly, the rate distortion function is
\[ R(D) = \frac{1}{2} \log \left( \frac{N_1 + N_2}{D} \right) > \frac{1}{2} \log \left( \frac{N_1 + \frac{PN_2}{P+N_2}}{D} \right).\]
\end{itemize}

\newpage
\section{Lattice coding for Compress-and-Forward}
\label{sec:LCF}

 \begin{wrapfigure}{r}{5cm}
\centering
\vspace{-0.4cm}
\includegraphics[width=5cm]{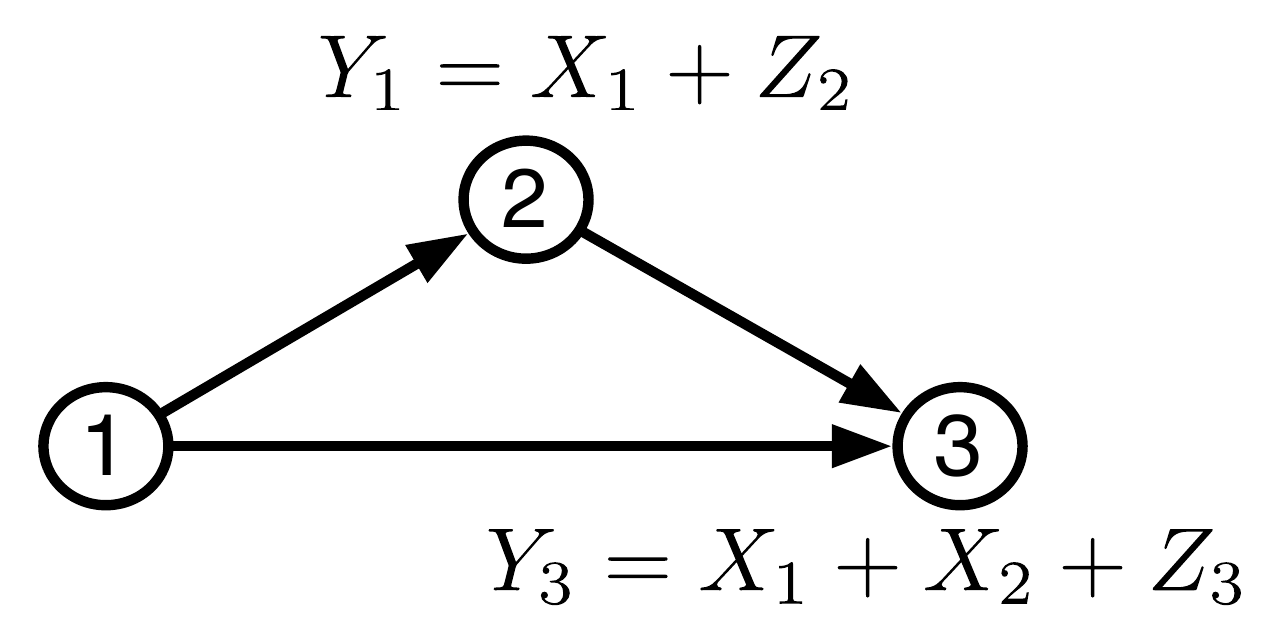}
\vspace{-0.9cm}
\caption{Three node Gaussian relay channel}
\label{fig:RC}
\end{wrapfigure} Using the general lattice-coding based Wyner-Ziv problem of the previous Section, we now implement a lattice Compress-and-Forward scheme for the classic three node Gaussian relay channel. The model is shown in Fig.\ref{fig:RC}, where the transmitter (Node 1) and the relay (Node 2) may transmit $X_1\in {\cal X}_1$ and $X_2\in {\cal X}_2$ subject to power constraints $E[|X_1|^2]\leq P_1$, $E[|X_2|^2]\leq P_2$, and $Y_2\in {\cal Y}_2$ and $Y_3\in {\cal Y}_3$ are the output random variables which are related to the inputs through the relationships in Fig. \ref{fig:RC}, where  $Z_2, Z_3$ are independent additive white Gaussian  noise of variance $N_2, N_3$. Furthermore, let $X_{i}(j)$ denote node $i$'s input at the $j$-th channel use, and let $X_1^n : = (X_1(1), X_1(2), \cdots, X_1(n))$.  Similar notation is used for received signals $Y_i(j)$. In this channel coding problem, we use classic definitions for achievable rates,  i.e. a $(2^{nR},n)$ code for a relay channel consists of a set of integers ${\cal W} = \{1,2, \cdots, 2^{nR}\}$, an encoding function $X_1: \{1,2,\cdots, 2^{nR}\} \rightarrow {\cal X}_1^n,$
a set of relay functions $\{f_i\}_{i=1}^n$such that $x_{2i} = f_i(Y_{2}(1), Y_{2}(2), \cdots, Y_{2}(i-1)), \;\;\;\; 1\leq i\leq n,$
and a decoding function $g: {\cal Y}_3^n \rightarrow \{1,2,\cdots 2^{nR}\}.$ We let the probability of error of this $(2^{nR},n)$ code be defined as $P_e^{(n)} : = \frac{1}{2^{nR}} \sum_{w \in {\cal W}} Pr\{g(Y_3^n) \neq w | w\mbox{ sent}\},$ for $w \in {\cal W}$.
The rate $R$ is said to be achievable if there exists a sequence of $(2^{nR},n)$ codes such that $P_e^{(n)}\rightarrow 0$ as $n\rightarrow \infty$.

 The Compress-and-Forward (CF) scheme originally proposed in  \cite{Cover:1979:relay} utilizes a  random coding argument, block Markov encoding, Wyner-Ziv
binning, and simultaneous joint typicality decoding. Our goal is to replace random codes with lattice codes and change the achievability techniques accordingly. In the CF scheme of \cite{Cover:1979:relay}, the relay compresses the received signal rather than decoding it, and transmits the bin index of its compression index. The destination first reconstructs the compressed signal the relay received using Wyner-Ziv coding, and then proceeds to decode the message from the combination of compressed signal received by the relay and the signal received by the destination itself.

\begin{thm}
For the three user Gaussian relay channel described by the input/output equations $Y_1 = X_1+Z_2$ and $Y_3 = X_1+X_2+Z_3$, with corresponding input and noise powers $P_1,P_2,N_2,N_3$, the following rate may be achieved using lattice codes in a lattice Compress-and-Forward fashion:
\begin{align*}
R & < \frac{1}{2} \log \left( 1 + \frac{P_1}{N_3} + \frac{P_1P_2}{P_1N_2 + P_1N_3 + P_2N_2 + N_2N_3} \right).
\end{align*}
\label{thm:LCF}
\end{thm}

The remainder of this Section is dedicated to the proof of Theorem \ref{thm:LCF}.

\medskip

{\bf Lattice codebook construction.} We employ three ``good'' lattice codebooks. Two of them are used as channel codebooks for the transmitter (Node 1) and the relay (Node 2). The third is used as a quantization/compression codebook by the relay. We drop all subscripts  / superscripts $n$ for ease of exposition and note that all lattices and lattice points are $n$-dimensional. %The codebooks are:%This means we need to use three pairs of 'good' nested lattice codes.\\
\begin{itemize}
\item Channel codebook for Node 1: codewords $t_1$ in codebook $\bold{C_1} = \{ \Lambda_{c1} \cap \mathcal{V}(\Lambda_1)\}$ where $\Lambda_1 \subseteq \Lambda_{c1} $ is a pair of good nested lattice codes --  $\Lambda_1$ is both Rogers-good and Poltyrev-good and $\Lambda_{c1}$ is Poltyrev-good.  We set $\sigma^2(\Lambda_1) = P_1$ to satisfy the transmitter  power constraint. We associate each message $w\in {\cal W}$ with the codeword $t_1$ in one-to-one fashion, $w \leftrightarrow t_1$, and send a dithered version of $t_1$. Note that $| \bold{C_1}| = 2^{nR}$.
\item Channel codebook for Node 2: codewords $t_2$ in codebook $\bold{C_2} = \{ \Lambda_{c2} \cap \mathcal{V}(\Lambda_2)\}$ where $\Lambda_2 \subseteq \Lambda_{c2} $ is a pair of good nested lattice codes: -- $\Lambda_2$ is both Rogers-good and Poltyrev-good and $\Lambda_{c2}$ is Poltyrev-good. We set $\sigma^2(\Lambda_2) = P_2$ to satisfy the relay power constraint. We associate each compression index $i$ with the codeword $t_2$ in one-to-one fashion: $i \leftrightarrow t_2$, and send a dithered version of $t_2$. Note that $| \bold{C_2}| = 2^{nR'} $.
\item Quantization/Compression codebook: $t_q \in \bold{C_q} = \{ \Lambda_{q} \cap \mathcal{V}(\Lambda)\}$ where $\Lambda \subseteq \Lambda_{q} $ is a pair of good nested lattice codes -- $\Lambda$ is Poltyrev-good and $\Lambda_q$ is Rogers-good.  We set $\sigma^2(\Lambda_q) = D$, $\sigma^2(\Lambda) = N_2 + \frac{P_1N_3}{P_1 + N_3} + D$, such that the source coding rate is $\widehat{R} = \frac{1}{2} \log \left( 1 + \frac{N_2 + \frac{P_1N_3}{P_1 + N_3}}{D} \right) $. These settings are explained in the following.
\end{itemize}

{\bf Encoding.} We use block Markov encoding as \cite{Cover:1979:relay}. In block $j$,  Node 1 chooses the codeword $t_1(j)$ associated with the message $w(j)$ to be transmitted in block $j$ and transmits
\[ X_1(j) = ( t_1(j) + U_1(j) ) \mod \Lambda_1, \]
where $U_1(j)$ is the dither signal which is uniformly distributed over $\mathcal{V}(\Lambda_1)$.
Node 2 quantizes the received signal in the last block $j-1$
\[ Y_2(j-1) = X_1(j-1) + Z_2(j-1) \]
 to $I(j-1)$ (with index $i(j-1)$) by using the quantization lattice code pair $(\Lambda_q, \Lambda)$ as described in the encoding part of Section \ref{sec:WZ}, where we set $\alpha_1 = 1$ and we set the second moment of $\Lambda$ to be $\sigma^2(\Lambda) = N_2 + \frac{P_1N_3}{P_1 + N_3} + D$. These settings will be explained later. Node 2 chooses the codeword $t_2(j-1)$ associated with the index $i(j-1)$ of $I(i-1)$ and sends
 \[X_2(j) = ( t_2(j-1) + U_2(j) ) \mod \Lambda \]
where $U_2$ is the dither signal which is uniformly distributed over $\mathcal{V} (\Lambda_2)$.

{\bf Decoding.} In block $j$, Node 3 receives
\[ Y_3(j) = X_1(j) + X_2(j) + Z_3(j). \]
It first decodes $t_2(j-1)$, and then the associated $I(j-1)$ and $X_2(j)$, using lattice decoding as in \cite{Erez:2004} subject to the channel coding rate constraint (recall that $t_2$ is of rate $R'$)
\[ R' < \frac{1}{2} \log \left( 1 + \frac{P_2}{P_1 + N_3} \right),\]
which ensures the correct decoding of $t_2(j-1)$.
We note that the source coding rate of $I$,
\[ \widehat{R} = \frac{1}{2} \log \left( 1 + \frac{N_2 + \frac{P_1N_3}{P_1 + N_3}}{D} \right), \]
must be less than the channel coding rate $R'$, which means
\begin{align}
 \frac{1}{2} \log \left( 1 + \frac{N_2 + \frac{P_1N_3}{P_1 + N_3}}{D} \right) <   \frac{1}{2} \log \left( 1 + \frac{P_2}{P_1 + N_3} \right). \label{D}
 \end{align}
Node 3  then subtracts the decoded $X_2(j)$ from $Y_3(j)$ and obtains
\begin{align*}
Y_3'(j) &= Y_3(j) - X_2(j)= X_1(j) + Z_3(j)
\end{align*}
which is used as direct-link side-information in the next block $j+1$.  In the previous block, Node 3 had also obtained $Y_3'(j-1) = X_1(j-1) + Z_3(j-1)$. Combining this with $I(j-1)$, Node 3 uses $Y_3'(j-1)$ as side-information to reconstruct $\widehat{Y}_2(j-1)$ as in the decoding part of Section \ref{sec:WZ}, with $\alpha_1 =1$, and $\sigma^2(\Lambda) = N_2 + \frac{P_1N_3}{P_1 + N_3} + D$.

Thus, we see that the CF scheme employs the $(X+Z_1,X+Z_2)$ Wyner-Ziv coding scheme of Section \ref{sec:WZ}
%From the above, we are able to make the following analogy with the general lattice Wyner-Ziv of Section \ref{sec:WZ}: the CF in three-node relay channel corresponds to general Wyner-Ziv coding in which
where the source to be compressed at the relay is $X_1 + Z_2$ and the side-information at the receiver (from the previous block) is $X_1+ Z_3$. One small difference from what was described in Section \ref{sec:WZ} is that $X_1$ is not strictly Gaussian distributed for finite $n$. However, $X_1$ will approach a Gaussian random variable as $n \rightarrow \infty$ since $\Lambda_1$ is Rogers-good.  The step
\[ P_{e, n} = Pr \{ ( \alpha_1( (1-\alpha_2) X_1 - \alpha_2 Z_3 + Z_2) - E_q ) \mod \Lambda \neq  \alpha_1( (1-\alpha_2) X_1 - \alpha_2 Z_3 + Z_2) - E_q  \} \]
of  \eqref{pe} in Section \ref{sec:WZ}  now corresponds to
\[ P_{e, n} = Pr \{ ( ( (1-\alpha_2) X_1 - \alpha_2 Z_3 + Z_2) - E_q ) \mod \Lambda \neq  ( (1-\alpha_2) X_1 - \alpha_2 Z_3 + Z_2) - E_q  \} \]
since we have  chosen $ \alpha_1 = 1$, since
\begin{align}
\frac{1}{n} &E || (1-\alpha_2) X_1 - \alpha_2 Z_3 + Z_2 - E_q ||^2 =   \frac{P_1N_3}{P_1 + N_3} + N_2 + D  = \sigma^2(\Lambda).\label{eq:a2}
\end{align}
Thus, the above error probability still goes to 0 as $n \rightarrow \infty$ since $X_1$, while not Gaussian in this case, may be treated as such as $n\rightarrow \infty$ as $\Lambda_1$ is Rogers-good. Essentially, $X_1$ may be treated just as $E_q$ is treated. 
%may also be treated as self noise $E_q$ since $\Lambda_1$ is also Rogers-good.  In the last section $X$ and $Z_1,Z_2$ are Gaussian, $E_q$ is the lattice random which is rogers-ggod. In this section, $X_1$ is also a rogers-good lattice random just as $E	_q$, that is why I say 'treated as $E_q$'. 
%\nrd{YIWEI: I don't get the subtleties of the above blue part -- why exactly must be take $\alpha_1=1$? What would happen otherwise? Oh, this step has nothing to do with $\alpha_1 =1$, I just rewrite it. Since $X_1$ is not exactly Gaussian, we need to say it approaches Gaussian and can be treated as so.$\alpha_1 = 1$ is used to give an independent quantization noise as shown next}
We also note that $\alpha_2$ is chosen so as to guarantee \eqref{eq:a2}. 
%\nrd{YIWEI: can you add exactly where the choices of $\alpha_1=1$ and $\alpha_2$ are needed? We need $\alpha_2$ because the equation added above. We need $\alpha_1 =1$ because of 'independent quantization noise' as shown below} 

The compressed $Y_2(j-1)$ may now be expressed as
\begin{align*}
\widehat{Y}_2(j-1) &= ( \alpha_1^2 - \alpha_1^2\alpha_2 + \alpha_2) X_1(j-1) + \alpha_2 ( 1 - \alpha_1^2) Z_3(j-1) + \alpha_1^2 Z_2(j-1) - \alpha_1 E_q(j-1) \\
&= X_1(j-1) + Z_2(j-1) - E_q(j-1)
\end{align*}
%by proper choice of $\alpha_2$(\nrd{this step does need a proper $\alpha_2$} ), 
where $E_q = (Y_2+U) \mod \Lambda$  (with $U$ the quantization dither which is uniformly distributed over $\Lambda$) is independent and uniformly distributed over $\mathcal{V}(\Lambda_q)$ with second moment $D$.
Now, Node 3 may decode $t_1(j-1)$ (and the associated $w(j-1)$ from $ Y_3'(j-1)$ and $\widehat{Y}_2(j-1)$ by first linearly and coherently combining them as
\begin{align*}
\frac{\sqrt{P_1}}{N_3}  Y_3'(j-1)  +\frac{\sqrt{P_1}}{N_2 + D} \widehat{Y}_2(j-1)
=&  \frac{\sqrt{P_1}}{N_3} \left( X_1(j-1) + Z_3(j-1)) + \frac{\sqrt{P_1}}{N_2 + D} ( X_1(j-1) + Z_2(j-1) - E_q(j-1) \right) \\
=& \left( \frac{\sqrt{P_1}}{N_3} + \frac{\sqrt{P_1}}{N_2 + D} \right) X_1(j-1) + \frac{\sqrt{P_1}}{N_3} Z_3(j-1) + \frac{\sqrt{P_1}}{N_2 + D} \left(Z_2(j-1) - E_q(j-1) \right).
\end{align*}
Since $E_q$ will approach a Gaussian random vector of variance $D$ as $n\rightarrow \infty$, the above equation may be treated as an AWGN  channel. Using modulo lattice decoding \cite{Erez:2004}, we may decode $t_1(j-1)$ (and the associated message $w(j-1)$) as long as
\[ R < \frac{1}{2} \log \left( 1 + \frac{P_1}{N_3} + \frac{P_1}{N_2 + D} \right).\]
Combining this with the constraint \eqref{D}, we obtain
\begin{align*}
R &< \frac{1}{2} \log \left( 1 + \frac{P_1}{N_3} + \frac{P_1}{N_2 + \frac{( N_2 + \frac{P_1N_3}{P_1 + N_3}) (P_1+N_3)}{P_2}} \right) = \frac{1}{2} \log \left( 1 + \frac{P_1}{N_3} + \frac{P_1P_2}{P_1N_2 + P_1N_3 + P_2N_2 + N_2N_3} \right),
\end{align*}
which is the CF rate achieved by Gaussian random codes on pg. 17-48 of \cite{Gammal:LN}.

\bigskip
{\bf Remarks:} Notice that there is a slight difference between the $(X+Z_1,X+Z_2)$ Wyner-Ziv coding scheme described in Section \ref{sec:WZ}  and its application to the Compress-and-Forward scheme for the three node Gaussian relay channel. The reason we choose $\alpha_1 = 1$ rather than optimal coefficient $\alpha_1  = \sqrt{1 - \frac{D}{N_2 + \frac{PN_3}{P+N_3}}}$, and $\sigma^2(\Lambda) = N_2 + \frac{P_1N_3}{P_1 + N_3} + D$ rather than $N_2 + \frac{P_1N_3}{P_1 + N_3}$ is because we would like the quantization/compression error $\widehat{Y}_2 - Y_2$ to be independent of all other terms, so that we may view $\widehat{Y}_2 = X_1 + N_2 - E_q$  as an equivalent AWGN channel. This convention is generally used in Gaussian compress-and-forward such as in \cite{Gammal:LN}. 
%\nrd{YIWEI: you explained the $\alpha=1$, but why the 2nd moment constraint? it is changed correspondingly. All these $\alpha$ tricks happen in the equation below equation (1).... I did not see this change correspondingly in this section (I added it to the previous section)....? where? Added now explained above}

\section{Conclusion}
\label{sec:conclusion}
We have demonstrated a lattice Compress-and-Forward scheme for the classic three user Gaussian relay channel. 
Given the structured nature of lattice codes, this provides an alternative, more practical, more geometric and intuitive understanding of CF in Gaussian networks. This lattice CF scheme opens the door to a more generic lattice-based achievability scheme for arbitrary networks, such as for example a structured version of the recent, general, noisy network coding scheme, or its combination with the recently introduced Compute-and-Forward framework. This is the subject of ongoing work.

\bibliographystyle{IEEEtran}
\bibliography{refs}
\end{document}